\documentclass[11pt]{article}
\usepackage{amssymb, amsmath}
\usepackage{graphics}
\usepackage[dvipdfmx]{graphicx}
\usepackage{eepic}
\usepackage{epic}
\usepackage{float}
\newtheorem{thm}{Theorem}[section]
\newtheorem{crl}[thm]{Corollary}

\newtheorem{lmm}[thm]{Lemma}

\newtheorem{prp}[thm]{Proposition}

\newtheorem{dfn}[thm]{Definition}
\newtheorem{rem}[thm]{Remark}
\def\bp{\text{\mathversion{bold}{$P$}}}
\def\bq{\text{\mathversion{bold}{$Q$}}}
\def\bm{\text{\mathversion{bold}{$M$}}}
\def\bv{\text{\mathversion{bold}{$v$}}}
\newcommand{\qed}{\hfill\hbox{\rule[-2pt]{6pt}{6pt}}}
\title{\Large\bf Addition in Jacobians of tropical hyperelliptic curves}
\author{Atsushi \textsc{Nobe}\footnote{e-mail: \texttt{nobe@faculty.chiba-u.jp}}\\[10pt]
Department of Mathematics, Faculty of Education, Chiba University,\\ 1-33 Yayoi-cho Inage-ku, Chiba 263-8522, Japan.}
\date{}
\begin{document}
%

\maketitle


\begin{abstract}      
We show that there exists a surjection from the set of effective divisors of degree $g$ on a tropical curve of genus $g$ to its Jacobian by using a tropical version of the Riemann-Roch theorem.
We then show that the restriction of the surjection is reduced to the bijection on an appropriate subset of the set of effective divisors of degree $g$ on the curve.
Thus the subset of effective divisors has the additive group structure induced from the Jacobian.
We finally realize the addition in Jacobian of a tropical hyperelliptic curve of genus $g$ via the intersection with a tropical curve of degree $3g/2$ or $3(g-1)/2$. 
\end{abstract}

\section{Introduction}
Let $P_0$ be a point on a tropical elliptic curve.
Suppose $P_1$ to be the point such that $P_1=P_0+T$, where $T$ is also a point on the tropical elliptic curve and ``+" is the addition in the curve \cite{Vigeland04,Nobe08}.
By applying the addition $+T$ repeatedly, we obtain the sequence of points $\{P_0,P_1,P_2,\ldots\}=\{P_0,P_0+T,P_0+2T,\ldots\}$ on the tropical elliptic curve. 
The sequence of points thus obtained can be regarded as a dynamical system on the curve and is often referred as the ultradiscrete QRT system \cite{Nobe08}.
Since each member of the ultradiscrete QRT system has the tropical elliptic curve as its invariant curve and its general solution can be given by using the ultradiscrete theta function \cite{TTGOR97,MZ06,Nobe08}, it is considered to be a two-dimensional integrable dynamical system.
It should be noted that the evolution of the ultradiscrete QRT system is given by a piecewise linear map.
On the other hand, if we suppose $P_1$ to be the point such that $P_1=P_0+P_0=2P_0$ and apply the duplication repeatedly then we obtain the sequence of points $\{P_0,P_1,P_2,\ldots\}=\{P_0,2P_0,4P_0,\ldots\}$ on the tropical elliptic curve as well. 
The sequence of points thus obtained can also be regarded as a dynamical system on the curve and is often referred as the solvable chaotic system \cite{KNT08,KKNT09}.
The solvable chaotic system also has the tropical elliptic curve as its invariant curve and its general solution can also be given by using the ultradiscrete theta function; nevertheless it can not be considered to be an integrable system because the inverse evolution is not uniquely determined.
Thus the additive group structure of the tropical elliptic curve leads to two kinds of dynamical systems, one is integrable and the other is not.

In analogy to the theory of plane curves over $\mathbb{C}$, there exists a family of tropical plane curves parametrized with an invariant called the genus of the curve.
A paradigmatic example of such a family of tropical plane curves consists of the tropical hyperelliptic curves.
Tropical elliptic curves are of course the members of the family labeled by the lowest genus.
Therefore, it is natural to consider that the additive group structure of the Jacobian of a tropical hyperelliptic curve also leads to several kinds of dynamical systems containing both an integrable one and a solvable chaotic one.
In this paper, in order to investigate a dynamical system arising from the additive group structure of the tropical hyperelliptic curve, we first review several important notions in tropical geometry to describe the Riemann-Roch theorem for tropical curves.
We also introduce a $(g+2)$-parameter family of tropical hyperelliptic curves of genus $g$ known as the set of the isospectral curves of $(g+1)$-periodic ultradiscrete Toda lattice. 
We then show the existence of a surjection between a tropical hyperelliptic curve and its Jacobian by using the tropical version of the Riemann-Roch theorem.
This surjection induces the group structure of an appropriate set of effective divisor of degree $g$ on the hyperelliptic curve from that of the Jacobian.
We further show that we can realize the addition in the Jacobian of the tropical hyperelliptic curve of genus $g$ as the addition of the $g$-tuples of points on the curve in terms of the intersection with a curve of genus 0.

\section{Riemann-Roch theorem for tropical curves}
We briefly review the notions of tropical curves as well as rational functions and divisors on them.
By using these tools we mention the Riemann-Roch theorem for tropical curves, which was independently found by Gathemann--Kerber \cite{GK06} and Mikhalkin--Zharkov \cite{MZ06} in 2006.
\begin{dfn}[Tropical curve]
A \textbf{metric graph} is a pair $(\Gamma,l)$ consisting of a graph $\Gamma$ together with a length function $l:E(\Gamma)\to\mathbb{R}_{>0}$, where $E(\Gamma)$ is the edge set of the graph $\Gamma$.
The first Betti number of $\Gamma$ is called the \textbf{genus} of $\Gamma$.
A \textbf{tropical curve} is a metric graph $(\Gamma,l)$ with a length function $l:E(\Gamma)\to\mathbb{R}_{>0}\cup\{\infty\}$,{\it  i.e.}, a metric graph with possibly unbounded edges.
\end{dfn}

\begin{dfn}[Divisor]
A \textbf{divisor} $D$ on a tropical curve $\Gamma$ is a formal $\mathbb{Z}$-linear combination of finite points on $\Gamma$
\begin{align*}
D=\sum_{P\in\Gamma} a_P P,
\end{align*}
where $a_P\in\mathbb{Z}$ and $a_P=0$ for all but finitely many $P\in\Gamma$.
\end{dfn}
The addition of two divisors $D=\sum_{P\in\Gamma} a_P P$ and $D^\prime=\sum_{P\in\Gamma} a^\prime_P P$ on a tropical curve $\Gamma$ are defined to be $D+D^\prime=\sum_{P\in\Gamma} (a_P+a_P^\prime) P$.
All divisors on $\Gamma$ then naturally compose an abelian group. 
We call it the \textbf{divisor group} of $\Gamma$ and denote it by ${\rm Div}(\Gamma)$. 
The \textbf{degree} $\deg D$ of a divisor $D=\sum_{P\in\Gamma} a_P P$ is defined to be the integer $\sum_{P\in\Gamma} a_P$.
The \textbf{support} ${\rm supp}\mkern2mu D$ of $D$ is defined to be the set of all points of $\Gamma$ occurring with a non-zero coefficient.
If all the coefficients $a_P$ of a divisor $D=\sum_{P\in\Gamma} a_P P$ are non-negative then the divisor is called \textbf{effective} and is written $D>0$.
We define the canonical divisor $K_{\Gamma}$ of $\Gamma$ to be
\begin{align*}
K_{\Gamma}
:=
\sum_{P\in\Gamma}\left({\rm val}(P)-2\right)P,
\end{align*}
where ${\rm val}(P)$ is the valence of the point $P\in\Gamma$ \cite{Z93}.
If $P$ is an inner point on an edge of $\Gamma$ then ${\rm val}(P)=2$, therefor such points never appear in $K_{\Gamma}$.
The canonical divisor $K_{\Gamma}$ of a tropical curve $\Gamma$ of genus $g$ is an effective divisor of degree $2g-2$ \cite{MZ06}.

\begin{dfn}[Rational function]
A \textbf{rational function} on a tropical curve $\Gamma$ is a continuous function $f:\Gamma\to\mathbb{R}\cup\{\pm\infty\}$ such that the restriction of $f$ to any edge of $\Gamma$ is a piecewise linear integral function. 
The values $\pm\infty$ can only be taken at the unbounded edges of $\Gamma$.
\end{dfn}
The \textbf{order} ${\rm ord}_P f$ of a rational function $f$ at $P\in\Gamma$ is defined to be the sum of the outgoing slopes of all segments of $\Gamma$ emanating from $P$.
If ${\rm ord}_P f>0$ then the point $P\in\Gamma$ is called the \textbf{zero} of $f$ of order ${\rm ord}_Pf$.
If ${\rm ord}_P f<0$ then $P$ is called the \textbf{pole} of $f$ of order $\left|{\rm ord}_Pf\right|$.
We define the \textbf{principal divisor} of a rational function $f$ (or divisor associated to $f$) on $\Gamma$ to be
\begin{align*}
(f)
:=
\sum_{P\in\Gamma}\left({\rm ord}_P f\right)P.
\end{align*}

\begin{rem}
For any rational function $f$ on a tropical curve $\Gamma$ we have
\begin{align*}
\deg (f)
=
\sum_{P\in\Gamma}\left({\rm ord}_P f\right)
=
0.
\end{align*}
\end{rem}

\begin{dfn}[Linear system]
Let $D$ be a divisor of degree $n$ on a tropical curve $\Gamma$.
We denote by $R(D)$ the set of all rational functions $f$ on $\Gamma$ such that the divisor $(f)+D$ is effective:
\begin{align*}
R(D)
:=
\left\{
f\ |\
(f)+D>0
\right\}.
\end{align*}
For any $f\in R(D)$ the divisor $(f)+D$ is a sum of exactly $\deg\left((f)+D\right)=\deg (f)+\deg D=n$ points by the above remark.
\end{dfn}

When formulating a statement about the dimensions of the linear systems $R(D)$ we have to be careful since $R(D)$ is in general not a vector space but a polyhedral complex and hence its dimension is ill-defined. 
The following definition serve as a replacement.

\begin{dfn}[Rank]
Let $D$ be a divisor of degree $n$ on a tropical curve $\Gamma$.
We define the rank $r(D)$ of the divisor $D$ to be the maximal integer $k$ such that for all choices of (not necessarily distinct) points $P_1,P_2,\ldots,P_k\in\Gamma$ we have $R(D-P_1-P_2-\cdots-P_k)\neq\emptyset$.
If $R(D)=\emptyset$ then we define $r(D)=-1$.
\end{dfn}
If we want to specify the curve $\Gamma$ in the notation of $R(D)$ and $r(D)$ we also write them as $R_\Gamma(D)$ and $r_\Gamma(D)$, respectively.

\begin{dfn}[Equivalence of divisors]
Two divisors $D$ and $D^\prime$ on a tropical curve $\Gamma$ are called \textbf{equivalent} and written $D\sim D^\prime$ if there exists a rational function $f$ on $\Gamma$ such that $D^\prime=D+(f)$.
\end{dfn}
If $D\sim D^\prime$ then $D^\prime=D+(f)$ for some rational function $f$ on $\Gamma$.
Then the map $R(D)\to R(D^\prime)$, $g\mapsto g+f$ is a bijection because we have $(g)+D^\prime=(g)+D+(f)=(g+f)+D$.
Thus we have $r(D)=r(D^\prime)$.
\begin{lmm}[See \cite{GK06}]\label{lmm:equiv}
Let $\bar\Gamma$ be a tropical curve and let $\Gamma$ be the metric graph obtained from $\bar\Gamma$ by removing all unbounded edges.
Then every divisor $D\in{\rm Div}(\bar\Gamma)$ is equivalent on $\bar\Gamma$ to a divisor $D^\prime$ with ${\rm supp}\mkern2mu D^\prime\subset\Gamma$.
If $D$ is effective then $D^\prime$ can be chosen to be effective as well.
Moreover, $R_\Gamma(D^\prime)\neq\emptyset$ if and only if $R_{\bar\Gamma}(D^\prime)\neq\emptyset$.
\qed
\end{lmm}
Let $\bar\Gamma$, $\Gamma$, and $D^\prime$ as in lemma \ref{lmm:equiv}.
By lemma \ref{lmm:equiv} any effective divisor on $\bar\Gamma$ is equivalent to a divisor of the same degree with the support on $\Gamma$.
Therefore we conclude that $r_{\bar\Gamma}(D^\prime)=r_\Gamma(D^\prime)$.

We then have the Riemann-Roch theorem.
\begin{thm}[Riemann-Roch theorem for tropical curves \cite{GK06,MZ06,BN07}]\label{thm:RRT}
For any divisor $D$ on a tropical curve $\bar\Gamma$ of genus $g$ we have
\begin{align*}
r(D)-r(K_{\bar\Gamma}-D)
=
\deg D+1-g.
\end{align*}
\qed
\end{thm}

It immediately follows a corollary of theorem \ref{thm:RRT}.
\begin{crl}\label{crl:CorRRT}
If $\deg D>2g-2$ then $r(D)=\deg D-g$.
\end{crl}

(Proof)\quad
Since $\deg K_{\bar\Gamma}=2g-2$, we have $\deg (K_{\bar\Gamma}-D)<0$.
This implies $R(K_{\bar\Gamma}-D)=\emptyset$ and hence $r(K_{\bar\Gamma}-D)=-1$.
By the Riemann-Roch theorem we have
\begin{align*}
r(D)
=
\deg D+1-g-1
=
\deg D-g
\end{align*}
as desired.
\qed

\section{Surjections between tropical curves and their Jacobians}
By applying the Riemann-Roch theorem to a tropical curve of genus $g$, one can obtain several propositions concerning a surjection from the set of effective divisors of degree $g$ of the tropical curve to its Jacobian.
Let $\bar\Gamma$ be a tropical curve of genus $g$.
We define $\mathcal{D}_0(\bar\Gamma)$ to be the subgroup of the divisor group ${\rm Div}(\bar\Gamma)$ of $\bar\Gamma$ generated by the divisors of degree 0.
We also define $\mathcal D_l(\bar\Gamma)$ to be the subgroup of $\mathcal D_0(\bar\Gamma)$ generated by the principal divisors\footnote{Note that any principal divisor is of degree 0.} of rational functions on $\bar\Gamma$.
\begin{dfn}[Picard group]
We define the Picard group 
${\rm Pic}^0(\bar\Gamma)$ of a tropical curve $\bar\Gamma$ to be the residue class group ${\rm Pic}^0(\bar\Gamma):=\mathcal D_0(\bar\Gamma)/\mathcal D_l(\bar\Gamma)$.
In particular, by lemma \ref{lmm:equiv} we have
\begin{align*}
{\rm Pic}^0(\bar\Gamma)
=
{\rm Pic}^0(\Gamma)
:=
\mathcal D_0(\Gamma)/\mathcal D_l(\Gamma).
\end{align*}
\end{dfn}

Let $\Omega(\Gamma)$ be the space of global 1-form on $\Gamma$.
Also let $\Omega(\Gamma)^\ast$ be the vector space of $\mathbb{R}$-valued linear functions on $\Omega(\Gamma)$.
Then the integral cycles $H_1(\Gamma,\mathbb{Z})$ form a lattice in $\Omega(\Gamma)^\ast$ by integrating over them.
\begin{dfn}[Jacobian]
We define the Jacobian $J(\Gamma)$ of a tropical curve $\bar\Gamma$ to be
\begin{align*}
J(\Gamma)
:=
\Omega(\Gamma)^\ast\slash
H_1(\Gamma,\mathbb{Z}).
\end{align*}
\end{dfn}

\begin{rem}
Since there exists an isomorphism between ${\rm Pic}^0(\Gamma)$ and $J(\Gamma)$ \cite{MZ06}, we can identify them:
\begin{align*}
J(\Gamma)
\simeq
{\rm Pic}^0(\Gamma)
=
\mathcal D_0(\Gamma)/\mathcal D_l(\Gamma).
\end{align*}
\end{rem}

We define the subset $\mathcal{D}_g^+(\Gamma)$ of ${\rm Div}(\bar\Gamma)$ to be
\begin{align*}
\mathcal{D}_g^+(\Gamma)
:=
\left\{
D\in{\rm Div}(\Gamma)\ |\
\deg D=g, D>0
\right\}.
\end{align*}

Fix an element $D^\ast$ of $\mathcal{D}_g^+(\Gamma)$.
Then we can define two maps $\phi: \mathcal{D}_g^+(\Gamma)\to\mathcal{D}_0(\Gamma)$ and $\bar\phi: \mathcal{D}_g^+(\Gamma)\to J(\Gamma)$
to be
\begin{align*}
\phi(A)=A-D^\ast
\qquad
\mbox{and}
\qquad
\bar\phi(A)\equiv A-D^\ast
\quad
\mbox{(mod $\mathcal D_l(\Gamma)$)},
\end{align*}
for $A\in\mathcal{D}_g^+(\Gamma)$, respectively.
We then have the following theorem.
\begin{thm}
The map $\bar\phi: \mathcal{D}_g^+(\Gamma)\to J(\Gamma)$ is surjective.
\end{thm}

(Proof)\quad
It is sufficient to show that there exists an element $A$ of $\mathcal{D}_g^+(\Gamma)$ such that $A-D^\ast\sim D$ for any $D\in\mathcal{D}_0(\Gamma)$.
By the Riemann-Roch theorem we have
\begin{align*}
r(D+D^\ast)
&=
\deg (D+D^\ast)+1-g+r(K_{\bar\Gamma}-D-D^\ast)\\
&=
g+1-g+r(K_{\bar\Gamma}-D-D^\ast)\\
&\geq0
\end{align*}
because $r(K_{\bar\Gamma}-D-D^\ast)\geq-1$ holds.
This means that $R(D+D^\ast)\neq\emptyset$ and hence there exists a rational function $h$ satisfying $(h)+D+D^\ast>0$.
Let $A=(h)+D+D^\ast$.
Then $\deg A=0+0+g=g$ and $A>0$ lead to $A\in\mathcal{D}_g^+(\Gamma)$.
Moreover we have
\begin{align*}
\bar\phi(A)
\equiv
(h)+D
\equiv
D
\quad
\mbox{(mod $\mathcal D_l(\Gamma)$)}.
\end{align*}
This completes the proof.
\qed

As in the non-tropical case, for a tropical curve of genus one the surjection $\bar\phi$ reduces to the bijection.
This fact was first found by Vigeland in 2004 \cite{Vigeland04}.
\begin{thm}
The map $\bar\phi: \mathcal{D}_g^+(\Gamma)\to J(\Gamma)$ is bijective if and only if $g=1$.
\end{thm}

(Proof)\quad
If we assume $\bar\phi(P)\equiv\bar\phi(Q)$ for some $P,Q\in\mathcal{D}_g^+(\Gamma)$ then we have $P-D^\ast\sim Q-D^\ast$ and hence $P-Q\sim0$.
This means that there exists a rational function $h$ satisfying $(h)=P-Q$.
Since $P$ is effective, $(h)+Q$ is effective as well.
Therefore the rational function $h$ is an element of $R(Q)$.
If we assume $g=1$ then we have $\deg Q=1>0=2g-2$.
By corollary \ref{crl:CorRRT} we obtain
\begin{align*}
r(Q)=\deg Q-g=1-1=0.
\end{align*}
This implies that $h\in R(Q)$ is a constant function, and hence $P-Q=(h)=0$.
Thus the map $\bar\phi$ is injective.

On the other hand, if we assume $g\geq2$ then we have $\deg Q=g\leq2g-2$ and hence $\deg(K_{\bar\Gamma}-Q)\geq0$.
Let $1$ be a constant function then $(1)+ K_{\bar\Gamma}-Q=K_{\bar\Gamma}-Q>0$.
This implies $1\in R(K_{\bar\Gamma}-Q)\neq\emptyset$ and hence $r(K_{\bar\Gamma}-Q)\geq0$.
Thus we have
\begin{align*}
r(Q)
=
\deg Q+1-g+r(K_{\bar\Gamma}-Q)
\geq
1.
\end{align*}
Therefore there exists a rational function $h$ satisfying $P-Q=(h)\neq0$.
\qed

Any element of $\mathcal{D}_g^+(\Gamma)$ is given by $P_1+P_2+\cdots+P_g$, where $P_1,P_2,\cdots,P_g$ are the points on the metric graph $\Gamma$ not necessarily distinct.
Let us denote the image of $P_1+P_2+\cdots+P_g\in\mathcal{D}_g^+(\Gamma)$ with respect to the map $\bar\phi$ by 
\begin{align*}
(P_1P_2\cdots P_g):=\bar\phi(P_1+P_2+\cdots+P_g).
\end{align*}
Since the map $\bar\phi: \mathcal{D}_g^+(\Gamma)\to J(\Gamma)$ is surjective, we have
\begin{align*}
J(\Gamma)
=
\left\{
(P_1P_2\cdots P_g)\ |\
P_1,P_2,\cdots,P_g\in\Gamma
\right\}.
\end{align*}
Thus the addition in $J(\Gamma)$ can be pulled back in $\Gamma$ by $\bar\phi$.
Actually, we define the addition of $g$-tuples $\bp=(P_1,P_2,\ldots,P_g)$ and $\bq=(Q_1,Q_2,\ldots,Q_g)$ of the points on $\Gamma$ to be
\begin{align*}
\bp
+
\bq
:=
(P_1P_2\cdots P_g)
+
(Q_1Q_2\cdots Q_g),
\end{align*}
where the addition in the right-hand-side is considered to be the one in $J(\Gamma)$.

\section{Tropical hyperelliptic curves}
In this and the subsequent sections we concentrate on a $(g+2)$-parameter family of smooth tropical curves of genus $g$ called the tropical hyperelliptic curves.

Let us consider the (non-tropical) hyperelliptic curve of genus $g$ defined by the polynomial of degree $2g+2$:
\begin{align*}
&f(x,y)
:=
y^2-P(x)^2+4c_{-1},\\
&
P(x)
:=
\sum_{i=0}^{g+1}c_i x^i,
\quad
c_{g+1}=1,\
c_i\in\mathbb{C}
\quad
\mbox{for $i=-1,0,\ldots,g$},
\end{align*}
where we assume that $f(x,0)=-P(x)^2+4c_{-1}$ has no multiple root.
By applying to $f(x,y)$ the birational transformation $(x,y)\mapsto (X,Y)$
\begin{align*}
X=x,\qquad
Y=\frac{1}{2}y-\frac{1}{2}P(x),
\end{align*}
we obtain the hyperelliptic curve given by the polynomial
\begin{align*}
F(X,Y)
:=
Y^2+YP(X)+c_{-1}.
\end{align*}

\begin{figure}[htb]
\centering
{\unitlength=.05in{\def\arraystretch{1.0}
\begin{picture}(80,85)(-50,-40)
\thicklines
\dottedline(-50,24)(30,24)
\dottedline(13,-38)(13,38)
\dottedline(-50,0)(30,0)
\thicklines
\put(-40,3){\line(1,0){15}}
\put(-40,-3){\line(1,0){15}}
\put(-25,3){\line(4,1){8}}
\put(-25,-3){\line(4,-1){8}}
\put(-25,3){\line(0,-1){6}}
\put(-17,5){\line(3,1){6}}
\put(-17,-5){\line(3,-1){6}}
\put(-17,5){\line(0,-1){10}}
\put(-11,7){\line(0,-1){14}}

\dashline{1}(-11,7)(-1,12)
\dashline{1}(-11,-7)(-1,-12)

\put(-1,12){\line(0,-1){24}}
\put(-1,12){\line(3,2){6}}
\put(-1,-12){\line(3,-2){6}}
\put(5,16){\line(0,-1){32}}
\put(5,16){\line(1,1){8}}
\put(5,-16){\line(1,-1){8}}
\put(13,24){\line(0,-1){48}}
\put(13,24){\line(1,2){4}}
\put(13,-24){\line(1,-2){4}}

\dashline{1}(-48,3)(-40,3)
\dashline{1}(-48,-3)(-40,-3)

\dashline{1}(17,32)(20,38)
\dashline{1}(17,-32)(20,-38)
\put(11,25){\makebox(0,0){$V_0$}}
\put(4,18){\makebox(0,0){$V_1$}}
\put(-2,14){\makebox(0,0){$V_2$}}
\put(-12,10){\makebox(0,0){$V_{g-2}$}}
\put(-18,8){\makebox(0,0){$V_{g-1}$}}
\put(-25,6){\makebox(0,0){$V_g$}}

\put(11,-26){\makebox(0,0){$V_0^\prime$}}
\put(4,-19){\makebox(0,0){$V_1^\prime$}}
\put(-2,-15){\makebox(0,0){$V_2^\prime$}}
\put(-12,-10){\makebox(0,0){$V_{g-2}^\prime$}}
\put(-18,-8){\makebox(0,0){$V_{g-1}^\prime$}}
\put(-25,-6){\makebox(0,0){$V_g^\prime$}}
\put(-47,-6){\makebox(0,0){$V_\infty^\prime$}}
\put(-47,6){\makebox(0,0){$V_\infty$}}
\put(24,37){\makebox(0,0){$V_{-\infty}$}}
\put(24,-37){\makebox(0,0){$V_{-\infty}^\prime$}}

\put(11,2){\makebox(0,0){$A_0$}}
\put(3,2){\makebox(0,0){$A_1$}}
\put(-3,2){\makebox(0,0){$A_2$}}
\put(-7.5,-1.5){\makebox(0,0){$A_{g-2}$}}
\put(-20.5,1.5){\makebox(0,0){$A_{g-1}$}}
\put(-27.5,1){\makebox(0,0){$A_g$}}

\put(27,-3){\makebox(0,0){$\displaystyle Y=\frac{c_{-1}}{2}$}}
\put(30,22){\makebox(0,0){$X$}}
\put(11,37){\makebox(0,0){$Y$}}
\put(15,22){\makebox(0,0){${O}$}}
\put(13,0){\circle{.8}}
\put(13,24){\circle*{.5}}
\put(13,-24){\circle*{.5}}
\put(5,16){\circle*{.5}}
\put(5,-16){\circle*{.5}}
\put(5,0){\circle{.8}}
\put(-1,12){\circle*{.5}}
\put(-1,-12){\circle*{.5}}
\put(-1,0){\circle{.8}}
\put(-11,7){\circle*{.5}}
\put(-11,-7){\circle*{.5}}
\put(-11,0){\circle{.8}}
\put(-17,5){\circle*{.5}}
\put(-17,-5){\circle*{.5}}
\put(-17,0){\circle{.8}}
\put(-25,3){\circle*{.5}}
\put(-25,-3){\circle*{.5}}
\put(-25,0){\circle{.8}}
\put(-48,-3){\circle*{.5}}
\put(-48,3){\circle*{.5}}
\put(20,38){\circle*{.5}}
\put(20,-38){\circle*{.5}}
\end{picture}
}}
\caption{
The tropical hyperelliptic curve $\bar\Gamma$.
}
\label{fig:THC}
\end{figure}
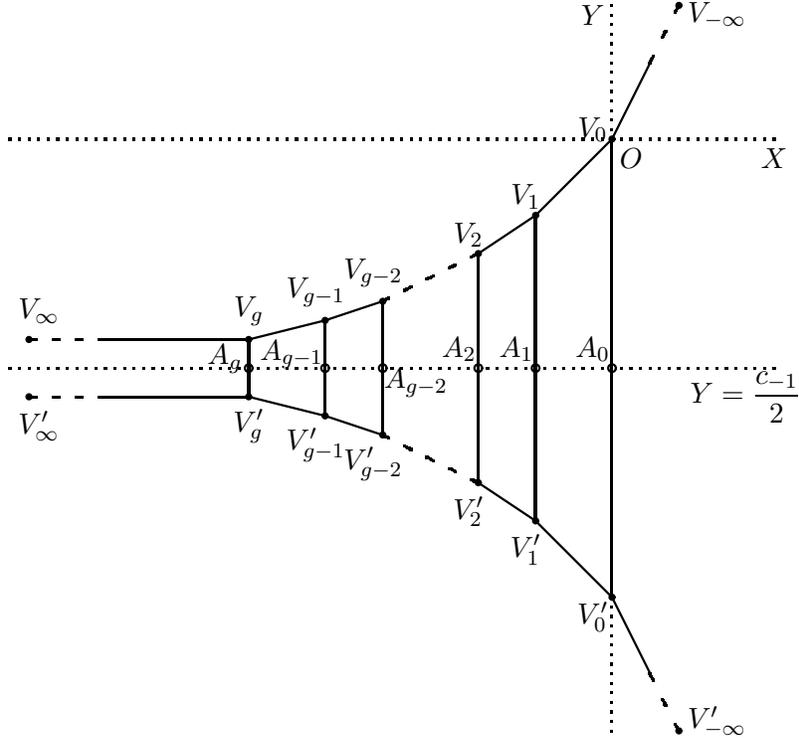

Now we tropicalize $F(X,Y)$. 
Replace the addition $+$ and the multiplication $\times$ in $F(X,Y)$ with the tropical addition $\oplus$ and the tropical multiplication $\otimes$, respectively.
If we define these tropical operations to be 
\begin{align*}
A\oplus B
:=
\max(A,B),
\qquad
A\otimes B
:=
A+B
\qquad
\mbox{for $A,B\in\mathbb{T}:=\mathbb{R}\cup\{-\infty\}$}
\end{align*}
then we obtain the following tropical polynomial
\begin{align*}
&\tilde{F}(X,Y)
:=
Y^{\otimes2}\oplus Y\otimes \tilde{P}(X)\oplus c_{-1}
=
\max\left(
2Y, Y+\tilde{P}(X),c_{-1}
\right),\\
&\tilde{P}(X)
:=
\bigoplus_{i=0}^{g+1}c_i\otimes X^{\otimes i}
=
\max_{i=0}^{g+1}\left(
c_i+i X
\right),
\end{align*}
where we assume $c_{g+1}=0$, $c_i\in\mathbb{T}$ for $i=-1,0,\ldots,g$.
Moreover assume
\begin{align*}
\begin{cases}
c_{g+1}=c_g=0,\\
c_{-1}<2c_0,\\
c_i+c_{i+2}<2c_{i+1}&\mbox{for $i=0,1,\ldots,g-2$},\\
c_{g-1}<2c_g=0.\\
\end{cases}
\end{align*}
Then the tropical polynomial $\tilde{F}(X,Y)$ defines a smooth tropical curve $\bar\Gamma$ of genus $g$ to be the set of its all non-differentiable points with respect to $X$ or $Y$ (see figure \ref{fig:THC}).

The curve $\bar\Gamma$ thus obtained is called the \textbf{tropical hyperelliptic curve}\footnote{A tropical curve is called hyperelliptic if there is a linear system with degree 2 and rank 1 \cite{HMY09}.} and is known as the isospectral curve of the ultradiscrete periodic Toda lattice, or the periodic box-ball system \cite{IT08}.

Let $\bv_e$ be the primitive tangent vector along the edge $e\in E(\bar\Gamma)$, where $E(\bar\Gamma)$ is the edge set of the tropical hyperelliptic curve $\bar\Gamma$.
Also let $|e|$ and $|\bv_e|$ be the Euclidean length of $e$ and $\bv_e$, respectively.
The length function $l:E(\bar\Gamma)\to\mathbb{R}_{>0}\cup\{\infty\}$ of $\bar\Gamma$ is defined to be
\begin{align*}
l(e)
:=
\frac{|e|}{|\bv_e|}.
\end{align*}

The vertex of $\bar\Gamma$ whose $Y$-coordinate is greater than $c_{-1}/2$  is denoted by $V_i$ and its conjugate (defined below) by $V_i^\prime$ ($i=0,1,\ldots,g$):
\begin{align*}
&V_0=(0,0),\quad V_0^\prime=(0,c_{-1}),\\
&V_i
=
\left(
c_{g-i}-c_{g-i+1}, (g-i+1)c_{g-i}-(g-i)c_{g-i+1}
\right)
&&
\mbox{for $i=1,2,\ldots,g$},\\
&V_i^\prime=
\left(
c_{g-i}-c_{g-i+1}, c_{-1}-(g-i+1)c_{g-i}+(g-i)c_{g-i+1}
\right)
&&
\mbox{for $i=1,2,\ldots,g$}.
\end{align*}
The outgoing slope of the edge $\overrightarrow{V_{i}V_{i+1}}$, which emanates from $V_i$ and connects it with $V_{i+1}$, is $g-i$ for $i=0,1,\ldots,g-1$.

We define the \textbf{bifurcation point }of the tropical hyperelliptic curve $\bar\Gamma$ to be the intersection point of $\bar\Gamma$ and the line $Y=c_{-1}/2$.
Note that $\bar\Gamma$ is symmetric under the refection with respect to $Y=c_{-1}/2$. 
There exist exactly $g+1$ bifurcation points on $\bar\Gamma$.
We denote the $g+1$ bifurcation points by $A_0,A_1,\ldots,A_g$ (see figure \ref{fig:THC}).
We also define the \textbf{conjugate} of a point $P$ on $\bar\Gamma$ to be the point which coincide with $P$ under the reflection with respect to $Y=c_{-1}/2$. 
The conjugate of a point $P$ on $\bar\Gamma$ is denoted by $P^\prime$. 
Thus the bifurcation point is characterized as the point $P$ on  $\bar\Gamma$ such that $P=P^\prime$.

The canonical divisors of $\bar\Gamma$ and $\Gamma$ are
\begin{align*}
&K_{\bar\Gamma}
=
\sum_{i=0}^g\left(V_i+V_i^\prime\right)-V_{-\infty}-V_{-\infty}^\prime-V_{\infty}-V_{\infty}^\prime,\\
&K_{\Gamma}
=
\sum_{i=1}^{g-1}\left(V_i+V_i^\prime\right),
\end{align*}
where $V_{-\infty}$, $V_{-\infty}^\prime$, $V_{\infty}$, and $V_{\infty}^\prime$ are the end points of the unbound edges (see figure \ref{fig:THC}).
Of course, $\deg K_{\bar\Gamma}=\deg K_{\Gamma}=2g-2$ holds.

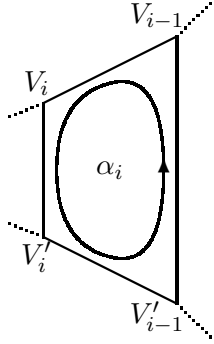
\begin{figure}[htbp]
\centering
{\unitlength=.035in{\def\arraystretch{1.0}
\begin{picture}(50,55)(-25,-25)
\thicklines
\dottedline(-10,10)(-15,8)
\dottedline(-10,-10)(-15,-8)
\dottedline(10,20)(15,25)
\dottedline(10,-20)(15,-25)
\thicklines

\put(-10,0){\line(0,1){10}}
\put(-10,0){\line(0,-1){10}}
\put(-10,10){\line(2,1){20}}
\put(-10,-10){\line(2,-1){20}}
\put(10,20){\line(0,-1){40}}

\qbezier(0,13)(-8,11)(-8,0)
\qbezier(-8,0)(-8,-11)(0,-13)
\qbezier(0,13)(8,15)(8,0)
\qbezier(0,-13)(8,-15)(8,0)
\put(8,0){\vector(0,1){2}}
\put(0,0){\makebox(0,0){$\alpha_i$}}
\put(-11,13){\makebox(0,0){$V_i$}}
\put(-11,-13){\makebox(0,0){$V_i^\prime$}}
\put(7,23){\makebox(0,0){$V_{i-1}$}}
\put(7,-22){\makebox(0,0){$V_{i-1}^\prime$}}
\end{picture}
}}
\caption{
The basis of the fundamental group $\pi_1(\Gamma)$.
}
\label{fig:Cycle}
\end{figure}

For the tropical hyperelliptic curve $\bar\Gamma$ we have a subset of $\mathcal{D}_{g}^{+}(\Gamma)$ on which the surjection $\bar\phi:\mathcal{D}_{g}^{+}(\Gamma)\to J(\Gamma)$ reduces to the bijection.
Let $\alpha_i$ be the basis of the fundamental group $\pi_1(\Gamma)$ of $\Gamma$ for $i=1,2,\ldots,g$ (see figure \ref{fig:Cycle}).
Also let $\alpha_{ij}=\alpha_i\cap\alpha_j\setminus\{\mbox{end points of $\alpha_i\cap\alpha_j$}\}$ for $i\neq j\in\{1,2,\ldots,g\}$.
We define the subset $\tilde{\mathcal{D}}_{g}^{+}(\Gamma)$ of $\mathcal{D}_{g}^{+}(\Gamma)$ to be
\begin{align*}
\tilde{\mathcal{D}}_{g}^{+}(\Gamma)
=
\left\{
\left.
\begin{array}{l}
P_1+\cdots+P_g\\
\end{array}
\right|\
\begin{array}{l}
\mbox{$P_i\in\alpha_i$ for all $i=1,2,\ldots,g$ and} \\
\mbox{there exists at most one point on $\alpha_{ij}$}
\end{array}
\right\}.
\end{align*}
We then have the following theorem.
\begin{thm}[See \cite{IT08}]\label{thm:bijection}
A reduced map $\bar\phi|_{\tilde{\mathcal{D}}_{g}^{+}(\Gamma)}$ is bijective
\begin{align*}
\bar\phi|_{\tilde{\mathcal{D}}_{g}^{+}(\Gamma)}:
\tilde{\mathcal{D}}_{g}^{+}(\Gamma)
\overset{\sim}{\to}
J(\Gamma).
\end{align*}
\qed
\end{thm}

We define the set $\Gamma_g^+$ to be the set of $g$-tuples $\bp=(P_1,P_2,\ldots,P_g)$ satisfying $P_1+P_2+\cdots+P_g\in\tilde{\mathcal{D}}_{g}^{+}(\Gamma)$:
\begin{align*}
\Gamma_g^+
:=
\left\{
\bp=(P_1,P_2,\ldots,P_g)\ |\ 
P_1+P_2+\cdots P_g\in\tilde{\mathcal{D}}_{g}^{+}(\Gamma)
\right\}.
\end{align*}
In terms of the map $\bar\phi$, the additive group structure of $\Gamma_g^+$ is induced from $J(\Gamma)$ .

\section{Realization of addition in tropical hyperelliptic curves}
\subsection{The case of even $g$}
Now we assume that the genus $g$ of the tropical hyperelliptic curve $\bar\Gamma$ is even.
Let us fix $D^\ast\in \mathcal{D}_g^+(\Gamma)$ as follows
\begin{align*}
D^\ast
=
\frac{g}{2}\left(V_0+V_0^\prime\right).
\end{align*}

\subsubsection{Unit of addition}

We have the following proposition.
\begin{prp}\label{prp:zero}
For any even $g$ we have
\begin{align*}
\left(P_1P_2\cdots P_{g/2}P_1^\prime P_2^\prime\cdots P_{g/2}^\prime\right)=0,
\end{align*}
where $P_1,P_2,\cdots, P_{g/2}$ are the points on $\Gamma$ and $P_1^\prime,P_2^\prime,\cdots, P_{g/2}^\prime$ are their conjugates, respectively.
\end{prp}

(Proof)\quad
Assume $(P_1P_2\cdots P_g)=0$.
Then we have
\begin{align*}
P_1+P_2+\cdots+P_g-D^\ast\sim0,
\end{align*}
and vice versa.
Therefore $(P_1P_2\cdots P_g)=0$ is equivalent to the existence of a rational function $h\in R(D^\ast)$ on $\Gamma$ such that $(h)=P_1+P_2+\cdots+P_g-D^\ast$.

\begin{figure}[htbp]
\centering
{\unitlength=.035in{\def\arraystretch{1.0}
\begin{picture}(50,55)(-25,-25)
\thicklines
\dottedline(-10,10)(-15,8)
\dottedline(-10,-10)(-15,-8)
\dottedline(10,-20)(10,20)
\dottedline(10,20)(15,25)
\dottedline(10,-20)(15,-25)
\thicklines
\put(-10,0){\line(0,1){10}}
\put(-10,0){\line(0,-1){10}}
\put(-10,10){\vector(2,1){20}}
\put(-10,-10){\vector(2,-1){20}}
\put(-13,0){\makebox(0,0){$A_i$}}
\put(-11,13){\makebox(0,0){$V_i$}}
\put(-11,-13){\makebox(0,0){$V_i^\prime$}}
\put(7,23){\makebox(0,0){$V_{i-1}$}}
\put(7,-22){\makebox(0,0){$V_{i-1}^\prime$}}
\put(0,12){\makebox(0,0){$\pi_i$}}
\put(0,-12){\makebox(0,0){$\pi_i^\prime$}}
\put(-10,0){\circle*{1}}
\put(10,20){\circle*{1}}
\put(10,-20){\circle*{1}}
\end{picture}
}}
\caption{
The paths $\pi_i$ and $\pi_i^\prime$.
}
\label{fig:path}
\end{figure}
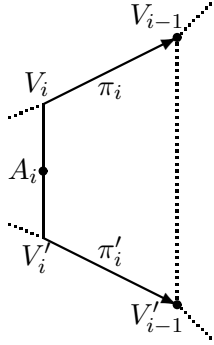

We show that there exists a rational function $h\in R(D^\ast)$ on $\Gamma$ such that $(h)=P_1+\cdots+P_{g/2}+P_1^\prime+\cdots+P_{g/2}^\prime-D^\ast$.
Without loss of generality we can assume the $Y$-coordinate of $P_i$ is greater than or equal to $c_{-1}/2$ for $i=1,2,\cdots,g/2$.
Let us denote the path in $\Gamma$ connecting $A_i$ with $V_{i-1}$ (resp. $V_{i-1}^\prime$) through $V_i$ (resp. $V_{i}^\prime$) by $\pi_i$ (resp. $\pi_{i}^\prime$) for $i=1,2,\ldots,g$ (see figure \ref{fig:path}).
Also denote the path in $\Gamma$ connecting $A_0$ with $V_0$ (resp. $V_{0}^\prime$) by $\pi_0$ (resp. $\pi_{0}^\prime$).

Owing to the above assumption, there exist $g/2$ points $P_1,P_2,\ldots,P_{g/2}$ on $\cup_{i=0}^g\pi_i$.
We construct the rational function $\bar h$ on $\cup_{i=0}^g\pi_i$ as follows.
The value of $\bar h$ at $A_g$ can arbitrarily be chosen.
We start walking the path $\pi_g$ form $A_g$ to $V_{g-1}$. 
If we find the point $P_{g_1}$ on the path $\pi_g$ then we set $\bar h(P_{g_1})=\bar h(A_g)$ and the outgoing slope of $\bar h$ from $P_{g_1}$ to be 1.
If we find the next point $P_{g_2}$ on the path $\pi_g$ then we set the outgoing slope of $\bar h$ from $P_{g_2}$ to be 2. 
Thus $P_{g_1}$ and $P_{g_2}$ are the zeros of $\bar h$ of order 1.
If $P_{g_1}=P_{g_2}$ then the part of $\bar h$ of slope 1 disappears and $P_{g_1}=P_{g_2}$ is the zero of $\bar h$ of order 2.
If there exist exactly $k$ points $P_{g_1},P_{g_2},\ldots,P_{g_k}$ on the path $\pi_g$ then the slope of $\bar h$ at $V_{g-1}$ is $k$.
The value of $\bar h$ at $V_{g-1}$ is uniquely determined.
By applying the same procedure inductively to the remaining paths $\pi_{g-1},\pi_{g-2},\ldots,\pi_0$ we obtain the rational function $\bar h$ on $\cup_{i=0}^g\pi_i$ satisfying 
\begin{align*}
(\bar h)=P_1+P_2+\cdots+P_{g/2}-\frac{g}{2}V_0.
\end{align*}
Note that the slope of $\bar h$ at $V_0$ is $g/2$ since there exist $g/2$ points $P_1,P_2,\ldots,P_{g/2}$ on $\cup_{i=0}^g\pi_i$.
Also note that the value of $\bar h$ at the bifurcation points $A_{g-1},A_{g-2},\ldots,A_0$ are uniquely determined by $h(V_{g-1}),h(V_{g-2}),\ldots,h(V_{0})$, respectively.

In the same manner we obtain the rational function $\bar h^\prime$ on $\cup_{i=0}^g\pi_i^\prime$ satisfying
\begin{align*}
(\bar h^\prime)=P_1^\prime+P_2^\prime+\cdots+P_{g/2}^\prime-\frac{g}{2}V_0^\prime.
\end{align*}
If we set $\bar h(A_g)=\bar h^\prime(A_g)$ then 
\begin{align*}
h=\bar h\cup \bar h^\prime
=
\begin{cases}
\bar h&\mbox{on $\cup_{i=0}^g\pi_i$},\\
\bar h^\prime&\mbox{on $\cup_{i=0}^g\pi_i^\prime$}\\
\end{cases}
\end{align*}
is  the rational function on $\Gamma=(\cup_{i=0}^g\pi_i)\cup(\cup_{i=0}^g\pi_i^\prime)$ satisfying $(h)=P_1+\cdots+P_{g/2}+P_1^\prime+\cdots+P_{g/2}^\prime-D^\ast$ as desired.
\qed

Combining theorem \ref{thm:bijection} and proposition \ref{prp:zero}, we obtain the following proposition concerning the unit of addition $\mathcal{O}$ of the group $(\Gamma_g^+,\mathcal{O})$.
\begin{prp}
For any even $g$, the $g$-tuple $(V_1,V_3,\cdots,V_{g-1},V_1^\prime,V_3^\prime,\cdots,V_{g-1}^\prime)$ is the unit of addition $\mathcal{O}$ of the group $(\Gamma_g^+,\mathcal{O})$:
\begin{align*}
\mathcal{O}
=
(V_1,V_3,\cdots,V_{g-1},V_1^\prime,V_3^\prime,\cdots,V_{g-1}^\prime).
\end{align*}
\end{prp}

(Proof)\quad
Since $V_i\in\alpha_i$ and $V_i^\prime\in\alpha_{i+1}$ and $V_i,V_i^\prime\not\in\alpha_{i,i+1}$ for $i=1,3,\ldots,g-1$, we have $(V_1,V_3,\cdots,V_{g-1},V_1^\prime,V_3^\prime,\cdots,V_{g-1}^\prime)\in\Gamma_g^+$. 
By proposition \ref{prp:zero} we have 
\begin{align*}
(V_1V_3\cdots V_{g-1}V_1^\prime V_3^\prime\cdots V_{g-1}^\prime)=0.
\end{align*}
On the other hand, $(V_1,V_3,\cdots,V_{g-1},V_1^\prime,V_3^\prime,\cdots,V_{g-1}^\prime)$ is the only choice which satisfies both $(V_1,V_3,\cdots,V_{g-1},V_1^\prime,V_3^\prime,\cdots,V_{g-1}^\prime)\in\Gamma_g^+$ and $(V_1V_3\cdots V_{g-1}V_1^\prime V_3^\prime\cdots V_{g-1}^\prime)=0$ because the map $\bar\phi|_{\tilde{\mathcal{D}}_{g}^{+}(\Gamma)}$ is bijective.
\qed

\subsubsection{Inverse elements}

For any element $\bp=(P_1,P_2,\ldots,P_g)$ of $\Gamma_g^+$ the inverse $-\bp$ with respect to the addition is simply given as follows.
\begin{prp}\label{prp:inv}
For any even $g$ we have
\begin{align*}
\bp
+
\bp^\prime
=
\mathcal{O},
\end{align*}
where  $\bp=(P_1,P_2,\ldots,P_g)\in\Gamma_g^+$ and $\bp^\prime:=(P_1^\prime,P_2^\prime,\ldots,P_g^\prime)$.
Therefore we write $-\bp=\bp^\prime$.
\end{prp}

(Proof)\quad
Note first that $\bp^\prime\in\Gamma_g^+$ since $\bp\in\Gamma_g^+$.
By proposition \ref{prp:zero} we have
\begin{align*}
\bp
+
\bp^\prime
&=
(P_1P_2\cdots P_g)
+
(P_1^\prime P_2^\prime\cdots P_g^\prime)\\
&\equiv
\sum_{i=1}^gP_i-D^\ast
+
\sum_{i=1}^gP_i^\prime-D^\ast
\quad
\mbox{(mod $J(\Gamma)$)}\\
&=
\sum_{i=1}^{g/2}(P_i+P_i^\prime)-D^\ast
+
\sum_{i=g/2+1}^g(P_i+P_i^\prime)-D^\ast\\
&\equiv
0\quad
\mbox{(mod $J(\Gamma)$)}\\
&=
\mathcal{O}.
\end{align*}
This completes the proof.
\qed

\subsubsection{Tropical curves passing through given points}

Now we realize the addition 
\begin{align}
\bp
+
\bq
=
\bm
\label{eq:add}
\end{align}
 of the group $(\Gamma_g^+,\mathcal{O})$, where $\bp=(P_1,P_2,\ldots,P_g)$, $\bq= (Q_1,Q_2,\ldots,Q_g)$, and $\bm=(M_1,M_2,\ldots,M_g)$ are the elements of $\Gamma_g^+$, by using the intersection of the tropical hyperelliptic curve $\bar\Gamma$ and a tropical curve of degree $3g/2$.
 
 By proposition \ref{prp:inv} the addition \eqref{eq:add} can be written
 \begin{align*}
\bp
+
\bq
+
\bm^\prime
=
\mathcal{O}.
\end{align*}
Therefore there exists a rational function $h$ on $\Gamma$ satisfying
\begin{align}
(h)
=
\sum_{i=1}^g\left(P_i+Q_i+M_i^\prime\right)-3D^\ast.
\label{eq:addrf}
\end{align}
This implies $h\in R(3D^\ast)$.

Define the rational functions $x$ and $y$ on the metric graph $\Gamma$ to be
\begin{align*}
x(P)=p,\qquad
y(P)=q,
\end{align*}
where $P=(p,q)$ is a point on $\Gamma$.
We can easily see
\begin{align*}
(x)=V_g+V_g^\prime-V_0-V_0^\prime,
\qquad
(y)=(g+1)V_0^\prime-(g+1)V_0.
\end{align*}
Consider the tropical monomial $a_i\otimes x^{\otimes i}$ in $x$ for $a_i\in\mathbb{T}$.
Then we have
\begin{align*}
a_i\otimes x^{\otimes i}
=
a_i+ix
\in R(3D^\ast)
\quad
\mbox{for $\displaystyle i=0,1,\ldots,\frac{3g}{2}$},
\end{align*}
where $a_0\otimes x^{\otimes 0}=a_0$ is the constant function.
Because the principal divisor is computed as follows
\begin{align*}
\left(
a_i+ix
\right)
=
\left(
ix
\right)
=
i\left(V_g+V_g^\prime\right)-i\left(V_0+V_0^\prime\right),
\end{align*}
and hence we have
\begin{align*}
\left(
a_i+ix
\right)
+3D^\ast
=
\left(
ix
\right)
=
i\left(V_g+V_g^\prime\right)+\left(\frac{3g}{2}-i\right)\left(V_0+V_0^\prime\right)>0
\end{align*}
for any even $g$ if and only if $i=0,1,\ldots,3g/2$.
Similarly, for the tropical monomial $b_i\otimes x^{\otimes i}\otimes y$ in $x, y$ for $b_i\in\mathbb{T} $ we have
\begin{align*}
b_i\otimes x^{\otimes i}\otimes y
=
b_i+ix+y
\in R(3D^\ast)
\quad
\mbox{for $\displaystyle i=0,1,\ldots,\frac{g}{2}-1$}.
\end{align*}
Because we have
\begin{align*}
\left(b_i+ix+y\right)
+3D^\ast
&=
\left(ix\right)
+
\left(y\right)
+3D^\ast\\
&=
i\left(V_g+V_g^\prime\right)
+
\left(\frac{5g}{2}+1-i\right)V_0^\prime
+
\left(\frac{g}{2}-1-i\right)V_0>0
\end{align*}
for any even $g$ if and only if $i=0,1,\ldots,g/2-1$.

Let us consider the tropical module $\mathcal{M}_g$ spanned by the $2g+1$ rational functions $x^{\otimes i}$ $(i=1,2,\ldots,3g/2)$ and $x^{\otimes i}\otimes y$ $(i=0,1,\ldots,g/2-1)$ \cite{MZ06}:
\begin{align*}
\mathcal{M}_g
:=&
\left\{
\bigoplus_{i=0}^{{3g}/{2}}a_{3g/2-i}\otimes x^{\otimes i}
\oplus
\bigoplus_{i=0}^{{g}/{2}-1}b_i\otimes x^{\otimes i}\otimes y
\ |\ 
a_i,b_i\in\mathbb{T}
\right\}\\
=&
\left\{
\max\left(
\max_{i=0}^{{3g}/{2}}\left(
a_{3g/2-i}+ix
\right),
\max_{i=0}^{{g}/{2}-1}\left(
b_i+ix+y
\right)
\right)\ |\ 
a_i,b_i\in\mathbb{T}
\right\}.
\end{align*}
Then we have the following proposition.
\begin{prp}
For any even $g$ we have
\begin{align*}
\mathcal{M}_g\subsetneq R(3D^\ast).
\end{align*}
\end{prp}

(Proof)\quad
Denote the set of poles of a rational function $h$ on the metric graph $\Gamma$ by $S_h$.
Then for the set $S_{f\oplus h}$ of the poles of the linear combination $f\oplus h=\max(f,h)$ of two rational functions $f$ and $h$ on $\Gamma$ we have
\begin{align*}
S_{f\oplus h}
\subset
S_f\cup S_h.
\end{align*}
Because if it does not hold then there exists a point $P\in S_{f\oplus h}$ which is the pole of neither $f$ nor $h$.
Since $P$ is the pole of $f\oplus h$, the sum of its outgoing slope at $P$ is negative, while those of both $f$ and $h$ are non-negative.
Let the edges outgoing from $P$ be $e_1,e_2,\ldots,e_k$ for $k=2,3$. ({Note that the point $P$ on $\Gamma$ is at most trivalent}.)
Also let the slope of $f$ and $h$ on $e_i$ be $f_i$ and $h_i$, respectively.
We then have $\sum_{i=1}^kf_i\geq0$ and $\sum_{i=1}^kh_i\geq0$.
By definition the outgoing slope of $f\oplus h$ at $P$ is
\begin{align*}
\sum_{i=1}^k\max\left(f_i,h_i\right)
\geq
\max\left(
\sum_{i=1}^kf_i,\sum_{i=1}^kh_i
\right)
\geq0.
\end{align*}
This is a contradiction.
Thus if $f,g\in R(3D^\ast)$ then $f\oplus h\in R(3D^\ast)$.
Therefore we have $\mathcal{M}_g\subset R(3D^\ast)$.

We can easily find a rational function $h\in R(3D^\ast)$ not included in $\mathcal{M}_g$, {\it e.g.}, the one whose two zeros are on the edge $\overrightarrow{V_0V_0^\prime}$. 
\qed

Let $h\in\mathcal{M}_g$.
Since $h$ is a tropical polynomial in $x$ and $y$ of degree $3g/2$, it can be extended as a rational function on the $(X,Y)$-plane.
If $P$ is the zero of $h$ then the function $h$ is not differentiable with respect to $X$ or $Y$ at $P$.
Thus the tropical curve $C$ of degree $3g/2$ defined by $h$ passes through the zero $P$ of $h$. 
If $P$ is on the tropical hyperelliptic curve $\bar\Gamma$ as well then $P$ is the intersection points of $\bar\Gamma$ and $C$.

\begin{figure}[htb]
\centering
{\unitlength=.035in{\def\arraystretch{1.0}
\begin{picture}(50,80)(-55,-30)
\thicklines
\put(-80,-20){\line(1,0){10}}
\put(-70,-20){\line(0,-1){10}}
\put(-70,-20){\line(5,1){10}}
\put(-60,-18){\line(0,-1){12}}
\dashline{2}(-60,-18)(-40,-10)
\put(-40,-10){\line(0,-1){20}}
\put(-40,-10){\line(2,1){6}}
\put(-34,-7){\line(0,-1){23}}
\put(-34,-7){\line(1,2){10}}
\put(-24,13){\line(0,1){32}}
\put(-24,13){\line(1,1){9}}
\put(-15,22){\line(0,1){23}}
\dashline{2}(-15,22)(-5,30)
\put(-5,30){\line(0,1){15}}
\put(-5,30){\line(5,3){10}}
\put(5,36){\line(0,1){9}}
\put(5,36){\line(5,2){10}}
\put(-36,-4){\makebox(0,0){$U_1$}}
\put(-41,-7){\makebox(0,0){$U_2$}}
\put(-57,-11){\makebox(0,0){$U_{3g/2-1}$}}
\put(-70,-16){\makebox(0,0){$U_{3g/2}$}}
\put(-20,11){\makebox(0,0){$U_0$}}
\put(14,32){\makebox(0,0){$U_{-g/2+2}$}}
\put(4,25){\makebox(0,0){$U_{-g/2+3}$}}
\put(-12,18){\makebox(0,0){$U_{-1}$}}
\end{picture}
}}
\caption{
The tropical curve $C$ of degree ${3g}/{2}$ defined by $h\in\mathcal{M}_g$.
}
\label{fig:TC}
\end{figure}
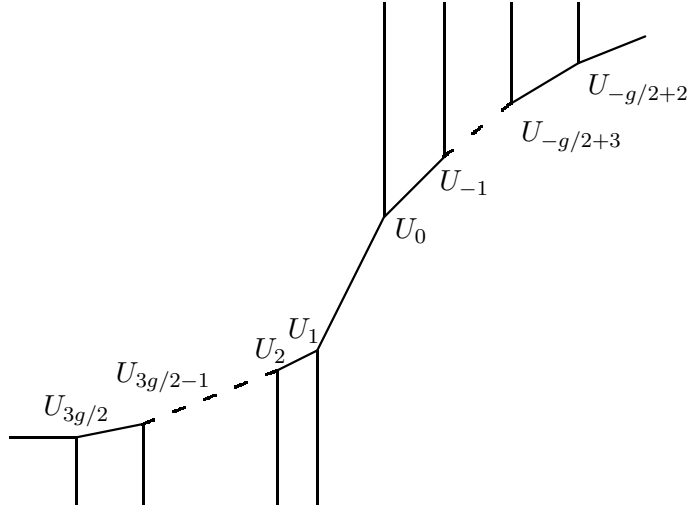

Assume that the coefficients of $h$ satisfy the generic condition
\begin{align}
\begin{cases}
a_{i}+a_{i+2}<2a_{i+1}
&
\mbox{for $\displaystyle i=0,1,\ldots,\frac{3g}{2}-2$},\\[5pt]
b_{i}+b_{i+2}<2b_{i+1}
&
\mbox{for $\displaystyle i=0,1,\ldots,\frac{g}{2}-3$},\\
a_1-a_0<b_0-b_1.
\end{cases}
\label{eq:gconde}
\end{align}
Then $C$ has exactly $2g-1$ vertices and is smooth.
Figure \ref{fig:TC} shows the smooth tropical curve $C$ satisfying the generic condition.

The vertex from which an unbound edge emanating upward is denoted by $U_i$ for non-positive $i=0,-1,\ldots,-g/2+2$; and the one from which an unbound edge emanating downward by $U_i$ for positive $i=1,2,\ldots,3g/2$.
The coordinate of the vertices are given as follows
\begin{align*}
&U_i
=
\left(
a_i-a_{i-1},\left(\frac{3g}{2}-i+1\right)a_i-\left(\frac{3g}{2}-i\right)a_{i-1}-b_0
\right)
&&
\mbox{for $i=1,2,\ldots,3g/2$},\\
&U_{-i}
=
\left(
b_i-b_{i+1},(i-1)b_{i+1}-ib_{i}+a_0
\right)
&&
\mbox{for $i=0,1,\ldots,g-2/2-1$}.
\end{align*}
The outgoing slope of the edge $\overrightarrow{U_iU_{i+1}}$ is $3g/2-i$ for $i=0,1,\ldots,3g/2-1$ and the one of the edge $\overrightarrow{U_{i-1}U_{i}}$ is $3g/2+i-1$ for $i=0,-1,\ldots, -(g-2)/2+2$.
The slopes of the remaining unbound edges emanating from $U_{-g/2+2}$ and $U_{3g/2}$ are $g+1$ and $0$, respectively.

\begin{rem}
If $h\in\mathcal{M}_g$ satisfies the generic condition \eqref{eq:gconde} then the curve $C$ defined by $h$ is smooth.
Therefore the dimension of the tropical module $\mathcal{M}_g$ is exactly $2g+1$ \cite{MZ06}.
Since $\deg 3D^\ast=3g>2g-2$ for any $g\geq2$, by corollary \ref{crl:CorRRT}, we have
\begin{align*}
r(3D^\ast)=3g-g=2g
=
\dim\mathcal{M}_g-1.
\end{align*}
This means that the maximal dimension of the cells of $R(3D^\ast)$ is at least $\dim\mathcal{M}_g$ \cite{GK06}.
\end{rem}

\subsubsection{Intersection numbers}

In order to investigate the intersection points of $\bar\Gamma$ and $C$, we consider the homogeneous coordinate $(x_0:x_1:x_2)$ of the tropical projective plane $\mathbb{TP}^{2}$ \cite{MZ06}.
Denote the point $(-\infty:0:-\infty)$ at infinity by $P_\infty$.
Then we can see $\bar\Gamma$ passes through $P_\infty$.
Actually, the unbound edge emanating from the vertex $V_0$ is given as follows
\begin{align}
(g+1)x_0-x_1-gx_2=0
\label{eq:pai}
\end{align}
in the homogeneous coordinate.
Note that this unbound edge goes to the direction $x_2=-\infty$.
If we set $x_2=-\infty$ in \eqref{eq:pai} then $x_0$ must be $-\infty$ since $g$ is positive.
We can also see that the $g/2-1$ unbound edges emanating from $U_{-i}$ for $i=0,1,\ldots,g/2-2$ of the curve $C$ pass through $P_\infty$.
Thus the two curves $\bar\Gamma$ and $C$ intersect at $P_\infty$.
We then have the following lemma concerning the intersection number of $\bar\Gamma$ and $C$ at $P_\infty$.
\begin{lmm}\label{lmm:evenint}
The tropical hyperelliptic curve $\bar\Gamma$ intersects the tropical curve $C$ defined by $h\in\mathcal{M}_g$ with multiplicity $3g^2/2$ at the point $P_\infty$ at infinity.
\end{lmm}

(Proof)\quad
Let the unbound edge of $\bar\Gamma$ emanating from the vertex $V_0$ be $e_0$.
Also let the edge of $C$ outgoing upward from the vertex $U_i$ be $\varepsilon_i$ for $i=0,-1,\ldots,-g/2+2$.
Let the remaining unbound edge of $C$ outgoing from $U_{-g/2+2}$ be $\varepsilon_{-g/2+1}$.
The primitive tangent vector of these unbound edges are given as follows
\begin{align*}
&e_0:\
\left(
g,g+1
\right),\\
&\varepsilon_i:\
\left(
1,1
\right)
\qquad
\mbox{for $i=0,-1,\ldots,-g/2+2$},\\
&\varepsilon_{-g/2+1}:\
\left(
g,g+1
\right),
\end{align*}
where we choose the basis $(-1,0)$ and $(1,1)$.

The intersection number at $P_\infty$ of two curves, whose primitive tangent vectors of the unbound edges passing through $P_\infty$ are $(a,b)$ and $(c,d)$ respectively, are defined to be 
\begin{align*}
w_1w_2
\min
\left(
ad, bc
\right),
\end{align*}
where $w_1$ and $w_2$ are the weights of the two edges, respectively \cite{Gathmann06}.
Note that the weights of the edges $e_0$ and $\varepsilon_i$ ($i=0,-1,\ldots,-g/2+1$) are all 1.
Then the intersection number of $\bar\Gamma$ and $C$ at $P_\infty$ can be computed as follows
\begin{align*}
\min
\left(
g(g+1),(g+1)g
\right)
+
\sum_{i=0}^{g/2-2}
\min
\left(
g\times1,(g+1)\times1
\right)
&=
g(g+1)
+
\left(\frac{g}{2}-1\right)g\\
&
=
\frac{3g^2}{2}.
\end{align*}
This completes the proof.
\qed

The degree of the curves $\bar\Gamma$ and $C$ are $g+2$ and ${3g}/{2}$, respectively.
By the B\'ezout theorem for tropical curves \cite{RST05,Gathmann06} $\bar\Gamma$ intersects $C$ at $(g+2){3g}/{2}$ points, counting multiplicities.
Note that $\bar\Gamma$ and $C$ do not intersect at the point at infinity other than $P_\infty$.
Therefore, in the affine part, $\bar\Gamma$ intersects $C$ at
\begin{align*}
\frac{3g}{2}(g+2)-\frac{3g^2}{2}
=
3g
\end{align*}
points, counting multiplicities.

\subsubsection{Realization of addition}

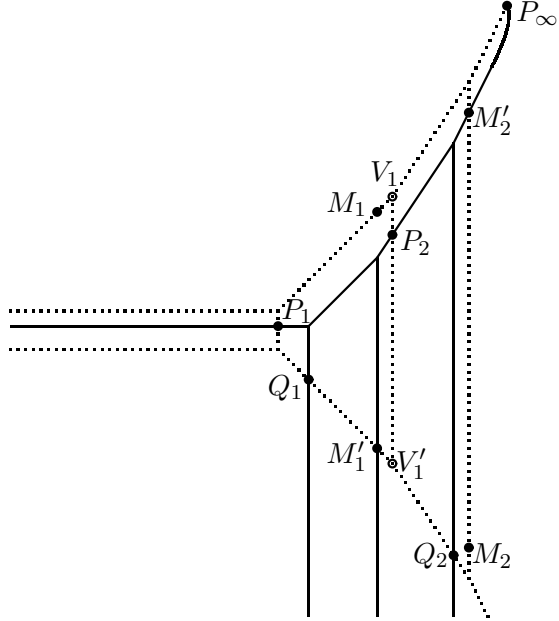
\begin{figure}[htb]
\centering
{\unitlength=.04in{\def\arraystretch{1.0}
\begin{picture}(70,85)(-35,-35)
\thicklines
\dottedline(-5,0)(-5,5)
\dottedline(-5,5)(10,20)
\dottedline(-5,0)(10,-15)
\dottedline(-5,5)(-40,5)
\dottedline(-5,0)(-40,0)
\dottedline(10,-15)(10,20)
\dottedline(10,20)(20,35)
\dottedline(10,-15)(20,-30)
\dottedline(20,-30)(20,35)
\dottedline(20,35)(25,45)
\dottedline(20,-30)(22.5,-35)
\thicklines
\put(-1,3){\line(-1,0){39}}
\put(-1,3){\line(1,1){9}}
\put(-1,3){\line(0,-1){38}}
\put(8,12){\line(2,3){10}}
\put(8,12){\line(0,-1){47}}
\put(18,27){\line(1,2){5}}
\put(18,27){\line(0,-1){62}}
\qbezier(23,37)(26,43)(25,45)
\put(-2.5,5){\makebox(0,0){$P_1$}}
\put(29,44){\makebox(0,0){$P_\infty$}}
\put(-4,-5){\makebox(0,0){$Q_1$}}
\put(4,-14){\makebox(0,0){$M_1^\prime$}}
\put(13,14){\makebox(0,0){$P_2$}}
\put(23,30){\makebox(0,0){$M_2^\prime$}}
\put(15,-27){\makebox(0,0){$Q_2$}}
\put(9,23){\makebox(0,0){$V_1$}}
\put(12.5,-15){\makebox(0,0){$V_1^\prime$}}
\put(4,19){\makebox(0,0){$M_1$}}
\put(23,-27){\makebox(0,0){$M_2$}}
\put(-5,3){\circle*{1.}}
\put(25,45){\circle*{1}}
\put(-1,-4){\circle*{1.}}
\put(8,-13){\circle*{1.}}
\put(8,18){\circle*{1.}}
\put(10,15){\circle*{1.}}
\put(18,-27){\circle*{1.}}
\put(20,31){\circle*{1.}}
\put(20,-26){\circle*{1.}}
\put(10,20){\circle{1.}}
\put(10,-15){\circle{1}}
\end{picture}
}}
\caption{
An intersection of the tropical hyperelliptic curve $\bar\Gamma$ (broken one) of genus 2 and the curve $C$ (solid one) of degree $3$.
The addition $(P_1,P_2)+(Q_1,Q_2)=(M_1,M_2)$ of the couples of points on $\bar\Gamma$ is realized by the intersection of $\bar\Gamma$ and $C$ with the unit $\mathcal{O}=(V_1,V_1^\prime)$.
}
\label{fig:g2int}
\end{figure}

Let $\bp=(P_1,P_2,\ldots,P_g)$ and $\bq= (Q_1,Q_2,\ldots,Q_g)$ be in $\Gamma_g^+$.
Assume the $2g$ points in $\bp$ and $\bq$ to be in generic position.
Then there exists a unique tropical curve $C$ passing through these $2g$ points and defined by a rational function $h\in\mathcal{M}_g$.
Note that the rational function $h\in\mathcal{M}_g$ is uniquely determined by the $2g$ points $\bp$ and $\bq$ up to a constant because $h$ has $2g+1$ parameters.
Consider the intersection of the tropical hyperelliptic curve $\bar\Gamma$ and the curve $C$.
Since $\bar\Gamma$ intersects $C$ at $3g$ points, there further exist $g$ intersection points, counting multiplicities.
Let these intersection points be $M_1^\prime,M_2^\prime,\ldots,M_g^\prime$.
Figure \ref{fig:g2int} shows an example of the intersection $\bar\Gamma$ and $C$ for $g=2$.

Then the principal divisor of the rational function $h$ defining $C$ has the form \eqref{eq:addrf}.
It follows that the $g$-tuples $\bp$, $\bq$, and $\bm=(M_1,M_2,\ldots,M_g)$ satisfy the addition formula 
\begin{align*}
\bp
+
\bq
=
\bm
\end{align*}
of the group $(\Gamma_g^+,\mathcal{O})$ (see figure \ref{fig:g2int}).

\subsection{The case of odd $g$}
Next we assume the genus $g$ of the tropical hyperelliptic curve $\bar\Gamma$ to be odd.
Let us fix $D^\ast\in \mathcal{D}_g^+(\Gamma)$ as follows
\begin{align*}
D^\ast
=
\frac{g-1}{2}\left(V_0+V_0^\prime\right)
+
V_0.
\end{align*}

\subsubsection{Unit of addition}

We have the following proposition in analogy to the case of even $g$.
\begin{prp}\label{prp:zeroodd}
For any odd $g$ we have
\begin{align*}
\left(V_0P_1P_2\cdots P_{(g-1)/2}P_1^\prime P_2^\prime\cdots P_{(g-1)/2}^\prime\right)=0,
\end{align*}
where $P_1,P_2,\cdots, P_{(g-1)/2}$ are the points on $\Gamma$ and $P_1^\prime,P_2^\prime,\cdots, P_{(g-1)/2}^\prime$ are their conjugates, respectively.
\end{prp}

(Proof)\quad
It is sufficient to show the existence of the rational function $h$ satisfying
\begin{align*}
(h)
&=
V_0+P_1+P_2+\cdots P_{(g-1)/2}+P_1^\prime+P_2^\prime+\cdots +P_{(g-1)/2}^\prime-D^\ast\\
&=
V_0+P_1+\cdots P_{(g-1)/2}+P_1^\prime+\cdots +P_{(g-1)/2}^\prime-\left(\frac{g-1}{2}\left(V_0+V_0^\prime\right)+V_0\right)\\
&=
P_1+P_2+\cdots +P_{(g-1)/2}+P_1^\prime+P_2^\prime+\cdots +P_{(g-1)/2}^\prime-\frac{g-1}{2}\left(V_0+V_0^\prime\right).
\end{align*}
It is obvious from proposition \ref{prp:zero}.
\qed

Combining theorem \ref{thm:bijection} and proposition \ref{prp:zeroodd}, it immediately follows the proposition concerning the unit $\mathcal{O}$ of addition the group $(\Gamma_g^+,\mathcal{O})$.
\begin{prp}
The $g$-tuple $(V_0,V_2,V_4,\cdots,V_{g-1},V_2^\prime,V_4^\prime,\cdots,V_{g-1}^\prime)$ is the unit of addition $\mathcal{O}$ of the group $(\Gamma_g^+,\mathcal{O})$:
\begin{align*}
\mathcal{O}
=
(V_0,V_2,V_4,\cdots,V_{g-1},V_2^\prime,V_4^\prime,\cdots,V_{g-1}^\prime).
\end{align*}
\qed
\end{prp}
 
\subsubsection{Addition in tropical elliptic curves}

Since the case of $g=1$ is distinctive, we first consider the case.
Let $g=1$ then $D^\ast=V_0$ and $\mathcal{O}=V_0$.
Note that the map $\bar\phi:\mathcal{D}_1^+(\Gamma)=\Gamma\to J(\Gamma)$ is bijective, and hence $\Gamma_1^+=\mathcal{D}_1^+(\Gamma)=\Gamma\simeq J(\Gamma)$.
For the rational functions $y$ and $x+y$ we have
\begin{align*}
&(y)+3D^\ast
=
2V_0^\prime+V_0>0,\\
&(x+y)+3D^\ast
=
V_1+V_1^\prime+V_0^\prime>0.
\end{align*}
Define the tropical module $\mathcal{M}_1$ to be
\begin{align*}
\mathcal{M}_1
=
\left\{
\max(a,b+y,c+x+y)\ |\ 
a,b,c\in\mathbb{T}
\right\}.
\end{align*}
It is easy to see that $\mathcal{M}_1= R(3D^\ast)$.

Let $h\in\mathcal{M}_1$.
Suppose $h$ to pass through the points $P,Q\in\Gamma_1^+=\Gamma\simeq J(\Gamma)$, which are in generic position.
Then $h$ is uniquely determined up to a constant.
The tropical curve $C$ defined by $h$ is of degree 2 and passes through the points $P_\infty$ and $P_\infty^\prime:=(0:-\infty:-\infty)$ at infinity.
The tropical elliptic curve $\bar\Gamma$ defined by the tropical polynomial 
\begin{align*}
\tilde F(X,Y)
=
\max\left(
2Y,Y+c_0,Y+X,Y+2X,c_{-1}
\right)
\end{align*}
is of degree 3 and passes through both $P_\infty$ and $P_\infty^\prime$ as well. 

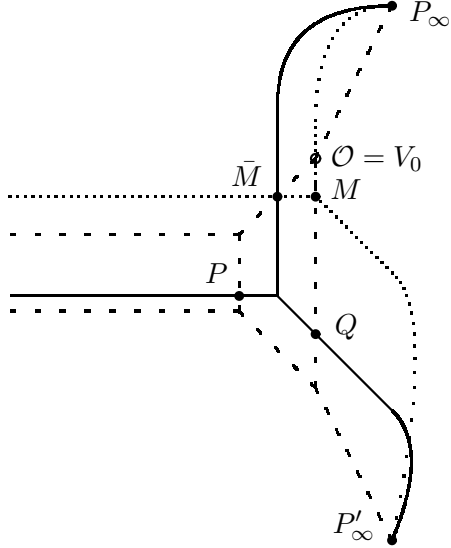
\begin{figure}[htb]
\centering
{\unitlength=.04in{\def\arraystretch{1.0}
\begin{picture}(50,75)(-40,-35)
\thicklines
\dashline{1}(-10,-5)(-10,5)
\dashline{1}(-10,5)(0,15)
\dashline{1}(-10,-5)(0,-15)
\dashline{1}(0,-15)(0,15)
\dashline{1}(0,15)(10,35)
\dashline{1}(0,-15)(10,-35)
\dashline{1}(-10,5)(-40,5)
\dashline{1}(-10,-5)(-40,-5)
\thicklines
\put(-40,-3){\line(1,0){35}}
\put(-5,-3){\line(1,-1){15}}
\put(-5,-3){\line(0,1){25}}
\qbezier(-5,22)(-5,35)(10,35)
\qbezier(10,-18)(15,-23)(10,-35)
\dottedline(-40,10)(0,10)
\dottedline(0,10)(0,20)
\dottedline(0,10)(10,0)
\qbezier[15](0,20)(0,35)(10,35)
\qbezier[20](10,0)(16,-5)(10,-35)
\put(-13,0){\makebox(0,0){$P$}}
\put(-9,13){\makebox(0,0){$\bar M$}}
\put(15,34){\makebox(0,0){$P_\infty$}}
\put(5,-33){\makebox(0,0){$P_\infty^\prime$}}
\put(8,15){\makebox(0,0){$\mathcal{O}=V_0$}}
\put(4,-7){\makebox(0,0){$Q$}}
\put(4,11){\makebox(0,0){$M$}}
\put(-10,-3){\circle*{1}}
\put(-5,10){\circle*{1}}
\put(0,-8){\circle*{1}}
\put(10,35){\circle*{1}}
\put(10,-35){\circle*{1}}
\put(0,10){\circle*{1}}
\put(0,15){\circle{1.2}}
\end{picture}
}}
\caption{
An intersection of the tropical elliptic curve $\bar\Gamma$ (broken one) and the curve $C$ (solid one) and $\tilde C$ (dotted one) of degree $2$.
The addition $P+Q=M$ of the points on $\bar\Gamma$ is realized by the intersection of $\bar\Gamma$ and $C$ and $\tilde C$ with the unit $\mathcal{O}=V_0$.
}
\label{fig:g1int}
\end{figure}

The primitive tangent vector of the unbound edge of $C$ passing through $P_\infty$ is $(1,1)$ with respect to the basis $(-1,0)$ and $(1,1)$; and the one of $\bar\Gamma$ passing through $P_\infty$ is $(1,2)$.
Therefore the intersection number of $C$ and $\bar\Gamma$ at $P_\infty$ is $\min(1\times2,1\times1)=1$.
The intersection number  of $C$ and $\bar\Gamma$ at $P_\infty^\prime$ can also be computed as $\min(2\times1,1\times3)=2$.
Therefore $\bar\Gamma$ intersects $C$ at 
\begin{align*}
2\times3-1-2=3
\end{align*}
points in its affine part, counting multiplicities.
Thus $\bar M\in\Gamma_1^+$ satisfying
\begin{align*}
P+Q+\bar M=\mathcal{O}
\end{align*}
is the third intersection point of $\bar\Gamma$ and $C$ (see figure \ref{fig:g1int}).

Let us consider the rational function $f\in\mathcal{M}_1$ passing through both $\bar M$ and $\mathcal O=V_0$.
Then the third intersection point $M$ (see figure \ref{fig:g1int}) of $\bar\Gamma$ and the curve $\tilde C$ defined by $f$ satisfies the addition
\begin{align*}
\bar M+M=\mathcal{O}.
\end{align*}
This implies that we have
\begin{align*}
P+Q=M.
\end{align*}
Thus the addition in $(\Gamma_1^+,\mathcal{O})$ can be realized as the intersection of $\bar\Gamma$ and tropical curves $C$ and $\tilde C$ of degree 2 (see figure \ref{fig:g1int}). 

\subsubsection{Tropical curves passing through given points}

Next we consider the case of $g\geq3$.
For the rational function $x$ we have
\begin{align*}
(ix)+3D^\ast
&=
i\left(V_g+V_g^\prime\right)+\left(\frac{3(g+1)}{2}-i\right)V_0+\left(\frac{3(g-1)}{2}-i\right)V_0^\prime>0
\end{align*}
for any odd $g\geq3$ if and only if $i=0,1,\ldots,3(g-1)/2$.
Moreover we have
\begin{align*}
(ix+y)+3D^\ast
&=
i\left(V_g+V_g^\prime\right)+\left(\frac{g+1}{2}-i\right)V_0+\left(\frac{5g-1}{2}-i\right)V_0^\prime
>0
\end{align*}
for any odd $g\geq3$ if and only if $i=0,1,\ldots,(g+1)/2$.

Define the tropical module $\mathcal{M}_g$ to be the linear combinations of the $2g+1$ rational functions $x^{\otimes i}$ $(i=0,1,\ldots,3(g-1)/2)$ and $x^{\otimes i}\otimes y$ $(i=0,1,\ldots,g+1/2)$:
\begin{align*}
\mathcal{M}_g
:=&
\left\{
\bigoplus_{i=0}^{{3(g-1)}/{2}}a_{3(g-1)/2-i}\otimes x^{\otimes i}
\oplus
\bigoplus_{i=0}^{g+1/{2}}b_i\otimes x^{\otimes i}\otimes y
\ |\ 
a_i,b_i\in\mathbb{T}
\right\}\\
=&
\left\{
\max\left(
\max_{i=0}^{{3(g-1)}/{2}}\left(
a_{3(g-1)/2-i}+ix
\right),
\max_{i=0}^{(g+1)/{2}}\left(
b_i+ix+y
\right)
\right)\ |\ 
a_i,b_i\in\mathbb{T}
\right\}.
\end{align*}
Then we have the following proposition in analogy to the case of even $g$.
\begin{prp}
For any odd $g$ we have
\begin{align*}
\mathcal{M}_g\subsetneq R(3D^\ast).
\end{align*}
\qed
\end{prp}

Assume that the rational function $h\in\mathcal{M}_g$ satisfies the generic condition
\begin{align*}
\begin{cases}
a_{i}+a_{i+2}<2a_{i+1}
&
\mbox{for $\displaystyle i=0,1,\ldots,\frac{3(g-1)}{2}-2$},\\[5pt]
b_{i}+b_{i+2}<2b_{i+1}
&
\mbox{for $\displaystyle i=0,1,\ldots,\frac{g+1}{2}-2$},\\
a_1-a_0<b_0-b_1.
\end{cases}
\end{align*}
Then the curve $C$ of degree $3(g-1)/2$ defined by $h$ has exactly $2g-1$ vertices and is smooth.

\begin{figure}[htbp]
\centering
{\unitlength=.035in{\def\arraystretch{1.0}
\begin{picture}(50,80)(-55,-30)
\thicklines
\put(-80,-20){\line(1,0){10}}
\put(-70,-20){\line(0,-1){10}}
\put(-70,-20){\line(5,1){10}}
\put(-60,-18){\line(0,-1){12}}
\dashline{2}(-60,-18)(-40,-10)
\put(-40,-10){\line(0,-1){20}}
\put(-40,-10){\line(2,1){6}}
\put(-34,-7){\line(0,-1){23}}
\put(-34,-7){\line(1,2){10}}
\put(-24,13){\line(0,1){32}}
\put(-24,13){\line(1,1){9}}
\put(-15,22){\line(0,1){23}}
\dashline{2}(-15,22)(-5,30)
\put(-5,30){\line(0,1){15}}
\put(-5,30){\line(5,3){10}}
\put(5,36){\line(0,1){9}}
\put(5,36){\line(5,2){10}}
\put(-36,-4){\makebox(0,0){$U_1$}}
\put(-41,-7){\makebox(0,0){$U_2$}}
\put(-57,-11){\makebox(0,0){$U_{3(g-1)/2-1}$}}
\put(-70,-15){\makebox(0,0){$U_{3(g-1)/2}$}}
\put(-20,11){\makebox(0,0){$U_0$}}
\put(14,32){\makebox(0,0){$U_{-(g+1)/2+1}$}}
\put(4,25){\makebox(0,0){$U_{-(g+1)/2+2}$}}
\put(-12,18){\makebox(0,0){$U_{-1}$}}
\end{picture}
}}
\caption{
The tropical curve $C$ of degree $3(g-1)/2$ defined by $h\in\mathcal{M}_g$.
}
\label{fig:TCodd}
\end{figure}
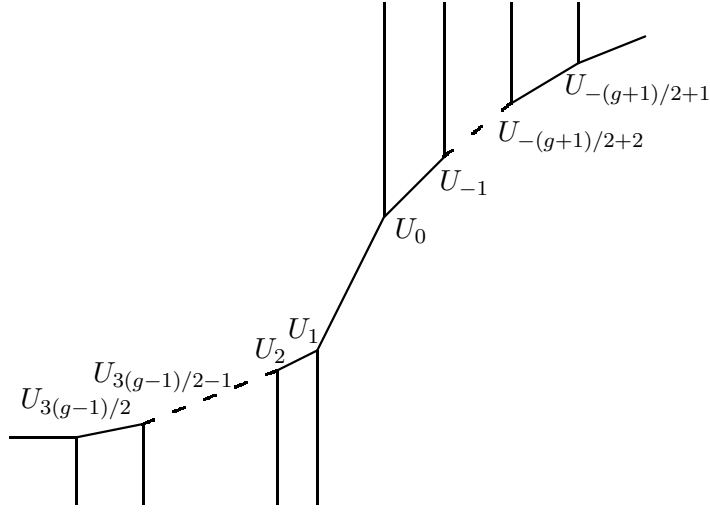

As in the case of even $g$, the vertex from which an unbound edge emanating upward is denoted by $U_i$ for non-positive $i=0,-1,\ldots,-(g+1)/2+1$; and the one from which an unbound edge emanating downward by $U_i$ for positive $i=1,2,\ldots,3(g-1)/2$.
The coordinates of the vertices are given as follows
\begin{align*}
&U_i
=
\left(
a_i-a_{i-1},\left(\frac{3(g-1)}{2}-i\right)\left(a_i-a_{i-1}\right)+a_i-b_0
\right)
&&
\mbox{for $i=1,2,\ldots,3(g-1)/2$},\\
&U_{-i}
=
\left(
b_i-b_{i+1},(i-1)b_{i+1}-ib_{i}+a_0
\right)
&&
\mbox{for $i=0,1,\ldots,(g+1)/2-2$}.
\end{align*}
The slope of the edge $\overrightarrow{U_iU_{i+1}}$ is $3(g-1)/2-i$ for $i=0,1,\ldots,3(g-1)/2-1$ and the one of the edge $\overrightarrow{U_{i-1}U_{i}}$ is $3(g-1)/2+i-1$ for $i=0,-1,\ldots, -(g+1)/2+2$.
The slopes of the remaining unbound edges emanating from $U_{-(g+1)/2+1}$ and $U_{3(g-1)/2}$ are $g-2$ and $0$, respectively (see figure \ref{fig:TCodd}).

\subsubsection{Intersection numbers}

We have the following lemma concerning the intersection number of $\bar\Gamma$ and $C$ at the point at infinity.
\begin{lmm}
For any odd $g\geq3$ the tropical hyperelliptic curve $\bar\Gamma$ of genus $g$ intersects the tropical curve $C$ defined by $h\in\mathcal{M}_g$ with multiplicity $3(g+1)(g-2)/2$ at the point $P_\infty$ at infinity.
\end{lmm}

(Proof)\quad
Let the unbound edge of $\bar\Gamma$ emanating from the vertex $V_0$ be $e_0$.
Also let the edge of $C$ outgoing upward from the vertex $U_i$ be $\varepsilon_i$ for $i=0,-1,\ldots,-(g+1)/2+1$.
Let the remaining unbound edge of $C$ outgoing from $U_{-(g+1)/2+1}$ be $\varepsilon_{-(g+1)/2}$.
The primitive tangent vector of these unbound edges are given as follows
\begin{align*}
&e_0:\
\left(
g,g+1
\right),\\
&\varepsilon_i:\ 
\left(
1,1
\right)
\qquad
\mbox{for $i=0,-1,\ldots,-(g+1)/2+1$},\\
&\varepsilon_{-(g+1)/2}:\
\left(
g-3,g-2
\right),
\end{align*}
where we choose the basis $(-1,0)$ and $(1,1)$.
The intersection number of $\bar\Gamma$ and $C$ at $P_\infty$ can be computed as follows
\begin{align*}
\min
\left(
g(g-2),(g+1)(g-3)
\right)
+
\sum_{i=0}^{g+1/2-1}
\min
\left(
g,g+1
\right)
&=
(g+1)(g-3)
+
\frac{g+1}{2}g\\
&
=
\frac{3(g+1)(g-2)}{2}.
\end{align*}
This completes the proof.
\qed

For any odd $g\geq3$ the degree of $\bar\Gamma$ and $C$ are $g+2$ and ${3(g-1)}/{2}$, respectively.
By the B\'ezout theorem $\bar\Gamma$ intersects $C$ at $3(g+2){(g-1)}/{2}$ points, counting multiplicities.
Note that $\bar\Gamma$ and $C$ do not intersect at the point at infinity other than $P_\infty$.
Therefore, in the affine part, $\bar\Gamma$ intersects $C$ at
\begin{align*}
\frac{3(g+2)(g-1)}{2}-\frac{3(g+1)(g-2)}{2}
=
3g
\end{align*}
points, counting multiplicities.

\begin{figure}[htb]
\centering
{\unitlength=.043in{\def\arraystretch{1.0}
\begin{picture}(90,85)(-45,-35)
\thicklines
\dottedline(-10,0)(-10,5)
\dottedline(-10,5)(0,10)
\dottedline(-10,0)(0,-5)
\dottedline(0,-5)(0,10)
\dottedline(0,10)(10,20)
\dottedline(0,-5)(10,-15)
\dottedline(-10,5)(-40,5)
\dottedline(-10,0)(-40,0)
\dottedline(10,-15)(10,20)
\dottedline(10,20)(20,35)
\dottedline(10,-15)(20,-30)
\dottedline(20,-30)(20,35)
\dottedline(20,35)(25,45)
\dottedline(20,-30)(22.5,-35)
\thicklines
\put(-40,3){\line(1,0){35}}
\put(-5,3){\line(2,1){8}}
\put(-5,3){\line(0,-1){38}}
\put(3,7){\line(1,1){5}}
\put(3,7){\line(0,-1){42}}
\put(8,12){\line(2,3){5}}
\put(8,12){\line(0,-1){47}}
\put(13,19.5){\line(1,1){4}}
\put(13,19.5){\line(0,1){15.5}}
\put(17,23.5){\line(2,1){26.5}}
\put(17,23.5){\line(0,1){11.5}}
\qbezier(13,35)(13,45)(25,45)
\qbezier(17,35)(17,45)(25,45)
\put(-11,7){\makebox(0,0){$P_1$}}
\put(-8,-4){\makebox(0,0){$Q_1$}}
\put(30,44){\makebox(0,0){$P_\infty$}}
\put(24,35){\makebox(0,0){$V_0$}}
\put(0,13){\makebox(0,0){$V_2$}}
\put(-1,-8){\makebox(0,0){$V_2^\prime$}}
\put(-2,2){\makebox(0,0){$\bar M_1$}}
\put(5,-6){\makebox(0,0){$P_2$}}
\put(6,-14){\makebox(0,0){$Q_2$}}
\put(13,13){\makebox(0,0){$\bar M_2$}}
\put(10,25){\makebox(0,0){$P_3$}}
\put(20,31){\makebox(0,0){$Q_3$}}
\put(23,23){\makebox(0,0){$\bar M_3$}}
\put(-10,3){\circle*{1.}}
\put(-5,-2.5){\circle*{1.}}
\put(0,5.5){\circle*{1.}}
\put(25,45){\circle*{1}}
\put(3,-8){\circle*{1.}}
\put(8,-13){\circle*{1.}}
\put(10,15){\circle*{1.}}
\put(13,24.5){\circle*{1.}}
\put(17,30.5){\circle*{1.}}
\put(20,25){\circle*{1.}}
\put(20,35){\circle{1.}}
\put(0,10){\circle{1}}
\put(0,-5){\circle{1.}}
\end{picture}
}}
\caption{
An intersection of the tropical hyperelliptic curve $\bar\Gamma$ (broken one) of genus 3 and the curve $C$ (solid one) of degree $3$. 
The addition $(P_1,P_2,P_3)+(Q_1,Q_2,Q_3)+(\bar M_1,\bar M_2,\bar M_3)=\mathcal{O}$ of the triples of points on $\bar\Gamma$ is realized by the intersection of $\bar\Gamma$ and $C$ with the unit $\mathcal{O}=(V_0,V_2,V_2^\prime)$.
}
\label{fig:g3int}
\end{figure}
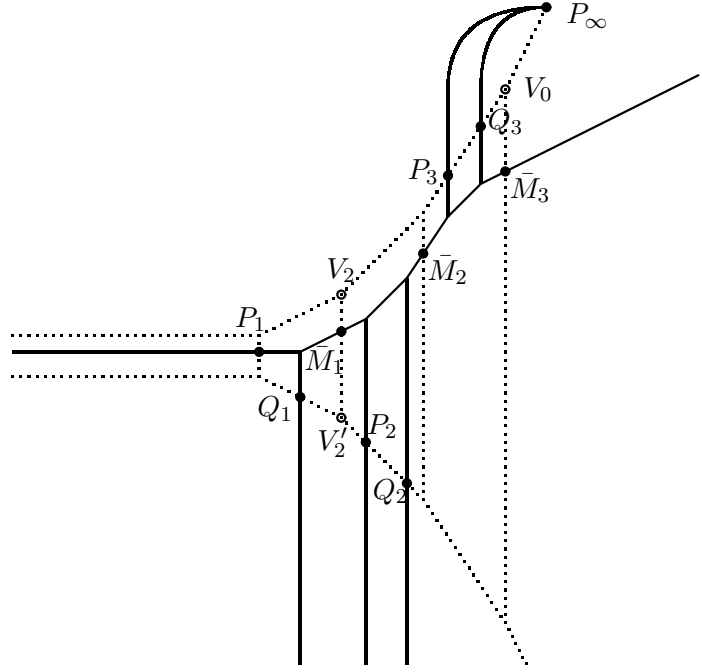

\subsubsection{Realization of addition}

Let $\bp=(P_1,P_2,\ldots,P_g)$ and $\bq= (Q_1,Q_2,\ldots,Q_g)$ be in $\Gamma_g^+$.
Assume the $2g$ points in $\bp$ and $\bq$ to be in generic position.
Then there exists a tropical curve $C$ passing through these $2g$ points and defined by a rational function $h\in\mathcal{M}_g$.
Consider the intersection of the tropical hyperelliptic curve $\bar\Gamma$ and the curve $C$.
Since $\bar\Gamma$ intersects $C$ at $3g$ points, there further exist $g$ intersection points, counting multiplicities.
Let these intersection points be $\bar M_1,\bar M_2,\ldots,\bar M_g$.
Then the principal divisor of the rational function $h$ defining $C$ has the form
\begin{align*}
(h)
=
\sum_{i=1}^g\left(P_i+Q_i+\bar M_i\right)-3D^\ast.
\end{align*}
Therefore we have
\begin{align*}
\bp
+
\bq
+
\bar\bm
=
\mathcal{O}.
\end{align*}
Figure \ref{fig:g3int} shows an example of the intersection of $\bar\Gamma$ of $g=3$ and $C$ of degree 3.

Moreover consider the curve $\tilde C$ defined by a rational function $f\in\mathcal{M}_g$ and passing through the $2g$ points in $\bar\bm=(\bar M_1,\bar M_2,\ldots,\bar M_g)$ and $\mathcal{O}$.
Then $\bar\Gamma$ intersects $\tilde C$ at $3g$ points, counting multiplicities.
Let the remaining $g$ intersection points be $M_1,M_2,\ldots,M_g$.
Then we have
\begin{align*}
\bm
+
\bar\bm
=
\mathcal{O},
\end{align*}
where $\bm=(M_1,M_2,\ldots,M_g)$.
It follows that the $g$-tuples $\bp$, $\bq$, and $\bm$ satisfy the addition formula 
\begin{align*}
\bp
+
\bq
=
\bm
\end{align*}
of the group $(\Gamma_g^+,\mathcal{O})$.

\section{Conclusion}
We show that there exists a surjection between the set of effective divisors of degree $g$ on the tropical hyperelliptic curve of genus $g$ and its Jacobian.
We also show that the surjection is bijective if and only if $g=1$.
We then show that there exists a subset of the set of effective divisors of degree $g$ on which the surjection reduces to the bijection.
It follows that the additive group structure of the subset is induced by the bijection from the Jacobian, which is isomorphic to the Picard group.
The addition in the Jacobian of a tropical hyperelliptic curve of genus $g$ thus induced can be interpreted geometrically as the addition of $g$-tuples of points on the curve.
We realize the addition of $g$-tuples of points on the curve in terms of the intersection of the hyperelliptic curve and a curve of degree $3g/2$ (resp. $3(g-1)/2$) for even (resp. odd) $g$. 

If $g=1$ then the addition of points on a tropical hyperelliptic curve induces two kinds of dynamical systems realized as the evolutions of points on the curve; one is an integrable system referred as the ultradiscrete QRT system and the other is a solvable chaotic system. 
Since we can realize the addition of $g$-tuples of points on the tropical hyperelliptic curve of genus $g$ via the intersection with a curve, we can construct several dynamical systems realized as the evolutions of points on the tropical hyperelliptic curve.
We will report such dynamical systems in a forthcoming paper.

\section*{Acknowledgments}
This work was partially supported by Grants-in-Aid for Scientific Research, Japan Society for the Promotion of Science (JSPS), No. 22740100.


\end{document}